\documentclass{ws-mpla}
\usepackage[super]{cite}
\usepackage{graphicx}
\begin{document}

\markboth{Giandomenico Palumbo}
{Gravitational Behaviour of an Effective Topological Field Theory in a Gravitational Instanton Background}

\catchline{}{}{}{}{}

\title{Gravitational Behaviour of an Effective Topological Field Theory in a Gravitational Instanton Background}

\author{\footnotesize Giandomenico Palumbo}

\address{School of Physics and Astronomy, University of Leeds, Leeds, LS2 9JT, United Kingdom\\
g.palumbo@leeds.ac.uk}

\maketitle

\pub{Received (Day Month Year)}{Revised (Day Month Year)}

\begin{abstract}
Effective topological field theories describe the properties of Dirac fermions in the low-energy regime. In this work, we introduce a new emergent gravity model by considering Dirac fermions invariant under local de Sitter transformations in four-dimensional open manifolds. In the context of Cartan geometry, fermions couple to spacetime through a $Spin(5)$ Cartan connection that can be decomposed in spin connection and tetrads. In presence of a gravitational instanton background, we show that the corresponding effective topological field theory becomes a dynamical gravitational theory with a positive cosmological constant and Barbero-Immirzi parameter. At classical level and in absence of matter, this theory is compatible with general relativity.

\end{abstract}

\ccode{PACS Nos.: 04.20.Gz, 04.62.+v, 04.20.-q}

\section{Introduction}
As is well known, general
relativity (GR)
describes gravity in the classical regime and characterizes the dynamical properties of spacetime. Intriguingly, in the first-order formalism, GR appears very similar to gauge theories and for this reason, in the MacDowell-Mansouri theory \cite{mac, wise}, it can be reformulated as a classical Yang-Mills-like theory, though it remains perturbatively non-renormalizable. 
Thus, if gravity is a fundamental force then the
problem of quantization should be solved by considering Einstein's
theory only as the low-energy limit of some more fundamental theory \cite{Polc}, or
alternatively, by considering the classical spacetime itself only a
rough approximation of a fundamental quantum spacetime \cite{Rovelli}. \\
However, there are alternative minimalist approaches \cite{Sindoni}, where gravity is
not considered a fundamental interaction and GR can be seen as a
low-energy effective theory induced by a standard quantum field
theory. This is the idea behind
Sakharov gravity \cite{sakh, visse}, where fermions, bosons, spacetime and metric
are already present at fundamental level and only the dynamics of spacetime,
defined as a Riemannian manifold, is emergent. Recently, similar results in presence of torsion were found in \cite{Stephan, Stephan2, Vassilevich} by employing the heat-kernel expansion in the context of Connes' spectral action principle \cite{Connes} which states that any physical action should be deducible from the spectral properties of some suitable Dirac operator.\\

The goal of this work is to overcome the two main problems of emergent gravity: firstly, in the Sakharov's approach there exist higher-order curvature terms in the effective action at one-loop that deviate from standard Einstein's theory and secondly, the value of cosmological constant, proportional to the Planck energy, is not compatible with the cosmological data related to accelerated expansion of the Universe. We avoid these problems by employing the first-order formalism in Cartan geometry (see, e.g. \cite{hehl, Sharpe, wise2, Westman}) with Lie algebra-valued forms \cite{wise2}. We start considering a Dirac fermions in open manifolds with Euclidean signature where they couple to the spacetime background through a $Spin(5)$ Cartan connection. In other words, following the gauge theory language, the corresponding spinor is considered invariant under local de Sitter transformations.
Integrating out the fermion fields we achieve the corresponding effective topological field theory in the low-energy limit which, in presence of a gravitational instanton background, becomes a gravitational theory with a positive cosmological constant and the Barbero-Immirzi parameter \cite{Immirzi} fixed to one \cite{Barbero}. We show that the corresponding equations of motion are compatible with the vacuum Einstein equations. 
Although the Einstein-Hilbert action is formally
equivalent to a constrained topological BF action \cite{Plebanski, baez} which is related to the MacDowell-Mansouri theory \cite{Freidel}, nevertheless we provide for the first time a derivation of the topological action from a fermion model at ground state choosing a suitable geometric framework.
Note that in the context of topological phases of matter, a link between topological theories, fermions and gravity was already highlighted in \cite{volo}.

\section{Dirac fermions in Cartan geometry and effective topological field theory}
We start considering massless
Dirac fermions living on four-dimensional open spin manifolds
$M$ with Euclidean signature. In the context of Cartan geometry, we take manifolds equipped with a de Sitter tangent space. This property reflects the fact that we are considering Dirac fermions invariant under local $Spin(5)$ transformations \cite{Randono2}, where the $Spin(5)$ group is the double covering of the $SO(5)$ de Sitter group. Thus, the corresponding fermion action is given by
\begin{eqnarray}
S[A,\overline{\psi},\psi]=-\frac{i}{\alpha^{2}}\int_{M} d^{4}x\,
|e|\;\overline{\psi}\,\displaystyle{\not}D^{A}\,\psi, \label{axialgrav}
\end{eqnarray}
where
\begin{eqnarray}\label{Dirac1}
\displaystyle{\not}D^{A}=\gamma^{j}e_{j}^{\mu}(\partial_{\mu}+A_{\mu}),
\end{eqnarray}
is the Dirac operator with Cartan connection $A_{\mu}$ that takes values in the $spin(5)$ algebra. Here $\mu=1,2,3,4$ is the spacetime index,  $\gamma^{j}$ are $4\times 4$ Euclidean Dirac matrices with $j=1,2,3,4,5$ \cite{Westman}, $\psi$ is a four-component spinor, $\alpha$ is a real dimensionless constant, $e_{\mu}=\frac{1}{2}\gamma_{j}e^{j}_{\mu}$ represent the tetrads (frame fields) and $|e|$ is their determinant. Clearly, in this framework, $A_{\mu}$ and $e_{\mu}$ are independent variables, implying the presence of a non-null torsion \cite{hehl2}. 
From this point until the end of the paper, in order to simplified the notation, we will employ the form-formalism \cite{wise2} saturating the spacetime index of the fields with the differential $dx^{\mu}$, i.e. $A=A_{\mu}dx^{\mu}$. 

We now focus on the physical properties of our fermion model in the low-energy regime. These are described by a
suitable effective field theory $S_\text{eff}[A]$ achieved by the fermion fields in the corresponding partition function of $S[A,\overline{\psi},\psi]$
\begin{eqnarray} \label{Dirac}
  e^{-S_{\text{eff}}[A]}\,=\,\int \mathcal{D}\,\overline{\psi}\; \mathcal{D}\psi\; e^{-S[A,\overline{\psi},\psi]}.
\end{eqnarray}
In general, any effective action can be decomposed in a topological $S_{\text{eff}}^{\text{top}}[A]$ and non-topological $S_{\text{eff}}^{\text{non-top}}[A]$ part, i.e. 
\begin{eqnarray}
S_{\text{eff}}[A]=S_{\text{eff}}^{\text{top}}[A]+S_{\text{eff}}^{\text{non-top}}[A], 
\end{eqnarray}
where the former is always the dominant one at ground state. Thus, in our case, the natural candidate for the effective topological action is simply given by
\begin{eqnarray}\label{bfgr}
S_{\text{eff}}^{\text{top}}[A]= -\frac{1}{192\, \pi\, \alpha^{2}}\int \text{tr}
\left(R_{A}\wedge R_{A}\right),
\end{eqnarray}
where $\text{tr}$ is the trace (Killing form) and $R_{A}=d_{A}A$ is the curvature form. This term is proportional to the topological Pontryagin invariant associated to $A$ and naturally generalizes the standard result in the Riemannian geometry framework \cite{Eguchi}.

Importantly, in Cartan geometry, the $Spin(5)$ connection can be decomposed in tetrads and spin connection $\omega_{\mu}$ \cite{wise, Randono2}. The latter takes values in $spin(4)$ which is the double covering of the $so(4)$ Lorentz algebra. The corresponding representation is given in terms of suitable products of the Dirac matrices $\gamma^{a}$ and $\gamma^{5}$ \cite{Randono2}, where $\gamma^{5}=i\gamma^{1}\gamma^{2}\gamma^{3}\gamma^{4}$ is the chiral matrix. Thus, we have that
\begin{eqnarray}
A=\omega + m\, e,
\end{eqnarray}
where $m$ is a dimensionful parameter and can be seen as a scale of energy. 
In this way, the curvature form $R_{A}$ decomposes in terms of $e$ and $\omega$, as follows
\begin{eqnarray}\label{A+Bgr}
R_{A}=R_{\omega}+m^{2}e\wedge e+m\, T_{\omega}.
\end{eqnarray}
Here, $R_{\omega}=d_{\omega}\omega$ and $T_{\omega}=d_{\omega}e$ are the curvature and torsion form of $\omega$, respectively.
Moreover, there exists an involutory automorphism $A\rightarrow \bar{A}$ with 
\begin{eqnarray}
\bar{A}=\omega-m\,e,
\end{eqnarray}
such that \cite{wise2}
\begin{eqnarray}\label{topo}
\text{tr}
\left(R_{A}\wedge R_{A}\right)=\text{tr}
\left(R_{\bar{A}}\wedge R_{\bar{A}}\right), \hspace{0.5cm}
\text{tr} \left(R_{A}\wedge \star R_{A}\right)=
-\text{tr}
\left(R_{\bar{A}}\wedge \star R_{\bar{A}}\right),
\end{eqnarray}
where $\star$ represents the internal Hodge dual ($\star^{2}=1$) and
\begin{eqnarray}\label{auto}
R_{\bar{A}}=R_{\omega}+m^{2}e\wedge e-m\, T_{\omega}.
\end{eqnarray}
We remark the fact that the above de Sitter connection $A$ plays a central role in the MacDowell-Mansouri theory where GR with a positive cosmological constant is reformulated as a Yang-Mills-like theory \cite{mac, wise}. In the next section, starting from the topological action (\ref{bfgr}), we derive a gravitational theory that coincides with the generalized MacDowell-Mansouri theory \cite{Randono, wise2}, where the Barbero-Immirzi parameter is naturally included.

\section{Emergent Euclidean gravity} 
We now show that in presence of a gravitational instanton background, the topological action (\ref{bfgr}) describes a dynamical gravitational theory.
In particular, the instanton background in the framework of Cartan geometry, can be defined through the following geometric conditions
\begin{eqnarray}\label{dualgr}
R_{A}=\star  R_{A}, \hspace{1.0cm} R_{\bar{A}}=-\star  R_{\bar{A}},
\end{eqnarray}
which are compatible with the relations (\ref{topo}).
Similar self-dual and anti-self dual relations appear in the context of gravitational instantons in Riemannian geometry \cite{Eguchi}. Thanks to (\ref{A+Bgr}) and (\ref{auto}), the above relations fix the following constraint between the spin connection and tetrads
\begin{eqnarray}\label{torsionnull}
m\, T_{\omega}=\star (R_{\omega}+m^{2}e\wedge e).
\end{eqnarray}
By employing (\ref{torsionnull}), we can replace the torsion in
(\ref{A+Bgr}) such that the effective action (\ref{bfgr}) becomes (at this point we omit the label ``top'')
\begin{eqnarray}\label{gravity}
S_{\text{eff}}[\omega, e]=-\frac{1}{96\, \pi\,
\alpha^{2}}\int \text{tr}\, \left[R_{\omega}\wedge \star R_{\omega} +R_{\omega}\wedge R_{\omega}+\right. m^{4}e\wedge e\wedge
e\wedge e+\nonumber \\ 2\,
m^{2}e\wedge e\wedge R_{\omega}+\left. 2\,m^{2}e\wedge e\wedge \star R_{\omega}+m^{4}e\wedge e\wedge
\star(e\wedge e)\right].
\end{eqnarray}
In this action only the last three terms are relevant at classical level simply because the first two terms are topological invariants with vanishing variation and the third one vanishes due to the graded Jacobi identity \cite{wise2} ($e\wedge e\wedge e=0$) while the fourth term is null only for the torsion-free connection.  By performing the following identifications
\begin{eqnarray}
m=\sqrt{\frac{\Lambda}{3}}, \hspace{1.0cm} \alpha=m\sqrt{\frac{
G}{3}},
\end{eqnarray}
where $\Lambda$ is a positive cosmological constant and $G$ is the
Newtonian constant,
the action (\ref{gravity}) acquires the following form
\begin{eqnarray}\label{gravitational-action}
S_{\text{eff}}[\omega, e]=-\frac{1}{16\pi G}\int\text{tr}\left[e\wedge e\wedge
\star R_{\omega}+\frac{\Lambda}{6}\;e\wedge e\wedge \star(e\wedge
e)+e\wedge e\wedge
R_{\omega}\right],
\end{eqnarray}
where we recognize the Hilbert-Palatini action with a cosmological constant while the third term represents the Holst term \cite{Holst, Baekler} with the Barbero-Immirzi parameter fixed to one \cite{Immirzi, Barbero}. Moreover, this action coincides with the generalized MacDowell-Mansouri action \cite{Randono, wise2}. It is important to remark the fact that although we started considering tetrads and spin connection as independent variables, because of the the constraint (\ref{torsionnull}), now we consider only the tetrads as free variables. By varying the action with respect to the tetrads, we have that
\begin{eqnarray}\label{TR}
 e\wedge R_{\omega}+e\wedge \star\left(R_{\omega}+\frac{\Lambda}{3}\;e\wedge
e\right)=d_{\omega}\,T_{\omega}+\sqrt{\frac{\Lambda}{3}}\,\,e \wedge T_{\omega}=d_{A}\,T_{\omega}=0,
\end{eqnarray}
where we have used $e\wedge R_{\omega}=d_{\omega}\,T_{\omega}$, i.e. the first Bianchi identity, and the constraint (\ref{torsionnull}). 
Moreover, by using the second Bianchi identity, i.e. 
\begin{eqnarray}
d_{A} R_{A}=0, \hspace{1.0cm} d_{A}=d_{\omega}+\sqrt{\frac{\Lambda}{3}}\,e\, \wedge, 
\end{eqnarray}
combined with (\ref{torsionnull}), (\ref{TR}), the graded Jacobi identity and the first Bianchi identity, we find that
\begin{eqnarray}\label{startorsion}
d_{A} R_{A}=d_{A}\left( R_{\omega}+\frac{\Lambda}{3}\, e\wedge e\right)+\sqrt{\frac{\Lambda}{3}}\,d_{A}\, T_{\omega}=\frac{\Lambda}{3}\, d_{\omega}(e\wedge e)-\frac{\Lambda}{3}\,\,e \wedge T_{\omega}=0.
\end{eqnarray}
From these equations and $T_{\omega}=d_{\omega} e$, the totally anti-symmetric part of the torsion (called also contorsion) $H$ must be null, namely
\begin{eqnarray}\label{dual}
H=e \wedge T_{\omega}=0,
\end{eqnarray}
which is the only component of the torsion that couples to fermions \cite{Obukhov}. 
This implies that $d_{\omega}T_{\omega}=0$ and $\omega$ reduces to the Levi-Civita connection $\widehat{\omega}$, i.e. $T_{\widehat{\omega}}=0$. 
In this way, the equations of motion (\ref{TR}) become
\begin{eqnarray}\label{Einstein}
 e\wedge \star\left(R_{\widehat{\omega}}+\frac{\Lambda}{3}\;e\wedge
e\right)=0,
\end{eqnarray}
which are the equations of GR with a positive cosmological constant and in absence of matter. Finally, we want to stress the self-consistency of the theory, pointing out that if the tetrads are invertible, as in the main physical cases, then the above equations imply 
\begin{eqnarray}
\star\left(R_{\widehat{\omega}}+\frac{\Lambda}{3}\;e\wedge
e\right)=0,
\end{eqnarray}
that coincides with the geometric constraint in (\ref{torsionnull}) in the case of Levi-Civita connection.

\section{Conclusions}
Summarizing, in this paper we have shown that a gravitational theory, compatible at classical level with Euclidean GR, arises from a
low-energy topological field theory induced by massless Dirac
fermions on curved spacetime with local de Sitter invariance. The two main ingredients employed in this derivation are the choice of Cartan geometry which naturally generalizes Riemannian geometry and the presence of a gravitational instanton background identified through self-dual and anti-self dual relations of the curvature tensor associated to the Cartan connection. This background implies the presence of a geometric constraint and the corresponding constrained topological field theory becomes a dynamical gravitational theory. This is not a real surprise because it is well known that a dynamically-trivial topological theory can acquire local propagating degrees of freedom in presence of constraints as shown for example in the BF-theory formulation of GR \cite{Plebanski, baez, Freidel}.\\
Importantly, in our approach, the cosmological constant
$\Lambda$ can assume, in principle, any positive value and is not related to the Planck energy scale as happens
in the standard emergent gravity scenario \cite{sakh, visse}. This property is desirable in emergent gravity, because only a small positive cosmological constant is compatible with the cosmological data related to accelerated expansion of the Universe. 
In conclusion, there are several open questions and challenges concerning the development of this theory on compact and non-compact Lorentzian spacetimes and in presence of matter. These points will be analyzed in a future work.

\end{document}